# Progressive field-state collapse and quantum non-demolition photon counting

Christine Guerlin[1], Julien Bernu[1], Samuel Deléglise[1], Clément Sayrin[1], Sébastien Gleyzes[1], Stefan Kuhr[1,†], Michel Brune[1], Jean-Michel Raimond[1] and Serge Haroche[1, 2]

**The irreversible evolution of a microscopic system under measurement is a central feature of quantum theory. From an initial state generally exhibiting quantum uncertainty in the measured observable, the system is projected into a state in which this observable becomes precisely known. Its value is random, with a probability determined by the initial system's state. The evolution induced by measurement (known as 'state collapse') can be progressive, accumulating the effects of elementary state changes. Here we report the observation of such a step-by-step collapse by measuring non-destructively the photon number of a field stored in a cavity. Atoms behaving as microscopic clocks cross the cavity successively. By measuring the light-induced alterations of the clock rate, information is progressively extracted, until the initially uncertain photon number converges to an integer. The suppression of the photon number spread is demonstrated by correlations between repeated measurements. The procedure illustrates all the postulates of quantum measurement (state collapse, statistical results and repeatability) and should facilitate studies of non-classical fields trapped in cavities.**

The projection of a microscopic system into an eigenstate of the measured observable reflects the change of knowledge produced by the measurement. Information is either acquired in a single step, as in a Stern–Gerlach spin-component measurement[1], or in an incremental way, as in spin-squeezing experiments[2,3]. A projective measurement is called 'quantum non-demolition'[4–8] (QND) when the collapsed state is invariant under the system's free unitary evolution. Sequences of repeated measurements then yield identical results and jumps between different outcomes reveal an external perturbation[7,8].

Various QND measurements have been realized on massive particles. The motional state of a trapped electron has been measured through the current induced in the trapping electrodes[9]. The internal state of trapped ions has been read out, directly by way of laser-induced fluorescence[10], or indirectly through quantum gate operations entangling them to an ancillary ion[11]. Collective spin states of an atomic ensemble have been QND-detected through its dispersive interaction with light[12].

QND light measurements are especially challenging, as photons are detected with photosensitive materials that usually absorb them. Photon demolition is however avoidable[13]. In non-resonant processes, light induces nonlinear dispersive effects[14] in a medium, without real transitions. Photons can then be detected without loss. Dispersive schemes have been applied to detect the fluctuations of a signal light beam by the phase shifts it induces on a probe beam interacting with the same medium[15,16]. Neither these methods, nor alternative ones based on the noiseless duplication of light by optical parametric amplifiers[17,18], have been able, so far, to pin down photon numbers.

Single-photon resolution requires an extremely strong light–matter coupling, optimally achieved by confining radiation inside a cavity. This is the domain of cavity quantum electrodynamics[19–21], in which experiments attaining single-quantum resolution have been performed with optical[22,23] or microwave photons, the latter being coupled either to Rydberg atoms[24–26] or to superconducting junctions[27]. In a QND experiment, cavity losses should be negligible during a sequence of repeated measurements. We have realized a superconducting cavity with a very long field damping time[28], and used it to detect repeatedly a single photon[29]. Here, we demonstrate with this cavity a general QND photon counting method applied to a microwave field containing several photons. It implements a variant of a procedure proposed in refs 30 and 31, and illustrates all the postulates of a projective measurement[1].

A stream of atoms crosses the cavity and performs a step-by-step measurement of the photon number-dependent alteration of the atomic transition frequency known as the 'light shift'. We follow the measurement-induced evolution from a coherent state of light into a Fock state of well-defined energy, containing up to 7 photons. Repeating the measurement on the collapsed state yields the same result, until cavity damping makes the photon number decrease. The measured field energy then decays by quantum jumps along a staircase-like cascade, ending in vacuum.

In this experiment, light is an object of investigation repeatedly interrogated by atoms. Its evolution under continuous non-destructive monitoring is directly accessible to measurement, making real the stochastic trajectories of quantum field Monte Carlo simulations[20,32]. Repeatedly counting photons in a cavity as marbles in a box opens novel perspectives for studying non-classical states of radiation.

## An atomic clock to count photons

To explain our QND method, consider the thought experiment sketched in Fig. 1a. A photon box, similar to the contraption imagined in another context by Einstein and Bohr[1], contains a few photons together with a clock whose rate is affected by the light. Depending upon the photon number $n$, the hand of the clock points in different directions after a given interaction time with the field. This time is set so that a photon causes a $\pi/q$ angular shift of the hand (here $q$ is an integer). There are $2q$ values (0, 1, …$2q-1$) of the photon number corresponding to regularly spaced directions of the hand, spanning 360° (Fig. 1a shows the hand's positions for $q = 4$ and $n = 0, 1, 3$). For $n \geq 2q$, the readings periodically repeat and the clock measures $n$ modulo $2q$.

[1] *Laboratoire Kastler Brossel, Département de Physique de l'Ecole Normale Supérieure, CNRS and Université Pierre et Marie Curie, 24 rue Lhomond, 75231 Paris Cedex 05, France*
[2] *Collège de France, 11 place Marcelin Berthelot, 75231 Paris Cedex 05, France*
[†] *Present address: Johannes Gutenberg Universität, Institut für Physik, Staudingerweg 7, 55128 Mainz, Germany*

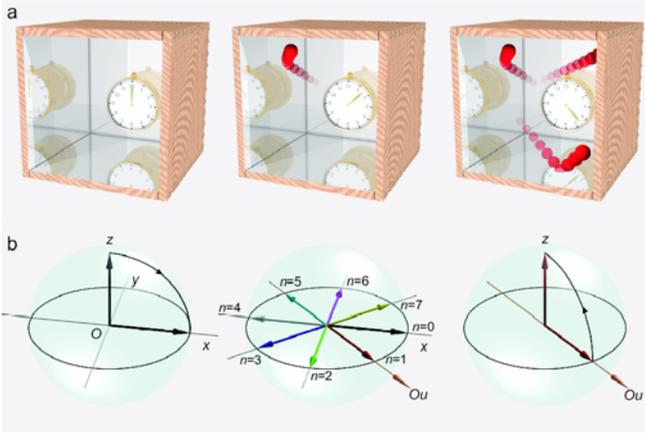

**Figure 1 Principle of QND photon counting. a**, Thought experiment with a clock in a box containing *n* photons. The hand of the clock undergoes a $\pi/4$ phase-advance per photon ($n = 0, 1, 3$ represented).
**b**, Evolution of the atomic spin on the Bloch sphere in a real experiment: an initial pulse $R_1$ rotates the spin from O–z to O–x (left). Light shift produces a $\pi/4$ phase shift per photon of the spin's precession in the equatorial plane. Directions associated with $n = 0$ to 7 end up regularly distributed over 360° (centre). Pulse $R_2$ maps the direction O–u onto O–z, before the atomic state is read out (right).

This description is translatable into the atomic world (Fig. 1b). The evolution of a two-level atom crossing the cavity C is described as the rotation of a spin evolving on a Bloch sphere[33]. The atomic levels, $|\pm_z\rangle$, correspond to up and down spin states along the O–z direction. Before entering C, the spin is rotated by a pulse $R_1$ from $|+_z\rangle$ into state $|+_x\rangle = (|+_z\rangle + |-_z\rangle)/\sqrt{2}$, represented by a vector along O–x. This vector then starts to rotate in the x–O–y equatorial plane, in analogy with the ticking of the hand in the thought experiment. The atomic flight time across C is adjusted to result, per photon, in a $\pi/q$ light-induced rotation of the spin. The $2q$ final spin states $|+_n\rangle$ correlated with $0 \leq n < 2q$ correspond to vectors reproducing the positions of the classical clock's hand.

In general, the photon number exhibits quantum uncertainty. The field, initially in a superposition $\sum_n c_n |n\rangle$ of Fock states $|n\rangle$, gets entangled with the spin, the final atom–field state becoming $\sum_n c_n |+_n\rangle \otimes |n\rangle$. The spin points in a fuzzy direction that the QND measurement is designed to pin down. As an example, consider a field in a coherent state[34] of complex amplitude $\alpha$ defined by the C-numbers $c_n = \exp(-|\alpha|^2/2)(\alpha^n/\sqrt{n!})$. Its photon number distribution, $P_0(n) = |c_n|^2$, is poissonian with an average photon number $n_0 = |\alpha|^2$ and a spread $\Delta n = \sqrt{n_0}$.

If we could determine the final atom state, the clock's delay— and hence $n$—would be read in a single measurement. This is however forbidden by quantum theory[35]. The $2q$ spin states are not mutually orthogonal (except for $q = 1$, see below) and cannot be unambiguously distinguished (this ambiguity is exploited in quantum cryptography[36]). Only partial information can be extracted from a spin, namely its projection along a direction O–u in the x–O–y plane making an arbitrary angle $\phi$ with O–x.

The angle between O–u and the direction of the $|+_n\rangle$ spin state is $n\pi/q - \phi$. The conditional probabilities for detecting the $|\pm_u\rangle$ states when C contains $n$ photons are $P(j,\phi|n) = [1 + \cos(n\pi/q - \phi + j\pi)]/2$ (using quantum information notation[33], we assign to the +/− spin states the values $j = 0/1$ and rename $|j,\phi\rangle$ the states $|\pm_u\rangle$). Measuring the spin along O–u is performed by submitting it, after cavity exit, to a pulse $R_2$ whose phase is set to map O–u onto O–z (Fig. 1b). This rotation is followed by the measurement of the atom's energy, equivalent to a spin detection along O–z. The combination of $R_1$ and $R_2$ is a Ramsey interferometer[37]. The probabilities for finding $j = 0$ and 1 along O–z oscillate versus $\phi$, which is a typical feature of quantum interference (Methods).

The $q = 1$ case is a notable exception for which a single measurement yields complete information. There are then only two opposite hand positions on the atomic clock dial, corresponding to orthogonal states. Ideally, a single detection pins down $n$ modulo 2, yielding the photon number parity. For weak fields with $n_0 \ll 1$, the probability for $n > 1$ is negligible and the parity defines $n$. The telegraphic signals[29] obtained by detecting a stream of atoms reveal the photon number evolution, with quantum jumps between $n = 0$ and 1 as the field randomly exchanges energy with the cavity walls. For larger fields, though, information must be extracted in a subtler way.

## Progressive pinning-down of photon number

The random outcome of a spin detection modifies our knowledge of the photon number distribution. The conditional probability $P(n|j,\phi)$ for finding $n$ photons after detecting the spin value $j$ along O–u is related to the inverse conditional probability $P(j,\phi|n)$ by Bayes' law[38]:

$$P(n|j,\phi) = P_0(n)P(j,\phi|n)/P(j,\phi) = P_0(n)[1 + \cos(n\pi/q - \phi + j\pi)]/2P(j,\phi) \quad (1)$$

where $P(j,\phi) = \sum_n P(j,\phi|n)P_0(n)$ is the a priori probability for $j$. This formula directly follows from the definition of conditional probabilities. It can also be derived from the projection postulate[1]. After detection of the spin in state $|j,\phi\rangle$, the entangled atom–field system collapses into $[\sum_n c_n \langle j,\phi|+_n\rangle |n\rangle] \otimes |j,\phi\rangle/\sqrt{P(j,\phi)}$. This entails that the photon number probability is (up to a global factor) multiplied by $|\langle j,\phi|+_n\rangle|^2 = P(j,\phi|n)$.

Equation (1) embodies the logic of our QND procedure. The spin measurement has the effect of multiplying $P_0(n)$ by $P(j,\phi|n)$, which is a periodic function vanishing for specific values of $n$ when $\phi$ is properly adjusted. If we choose $\phi = \pi p/q$ (where $p$ is an integer), O–u points along the direction of $|+_p\rangle$. This entails $P(j = 1, \phi|p) = P(j = 0, \phi|p + q) = 0$. Detecting the spin in 1 (resp. 0) excludes the photon number $n = p$ (resp. $p+q$) as these outcomes are forbidden for the corresponding Fock states. One of the probabilities for finding $n = p$ or $n = p + q$ is cancelled, while the other is enhanced (when normalization is accounted for). At the same time, the probabilities of other photon numbers are modified according to equation (1).

This decimation is robust against imperfections. In a realistic situation, the theoretical probability $P(j,\phi|n)$ becomes $P^{(\exp)}(j,\phi|n) = [A + B\cos(n\Phi - \phi + j\pi)]/2$, where $A$ and $B$ are the Ramsey interferometer fringes offset and contrast, somewhat different from 1. The phase shift per photon $\Phi$ may also slightly depart from $\pi/q$. Before a QND measurement, $A$, $B$ and $\Phi$ are determined by independent calibration. The limited contrast of the interferometer corresponds to a statistical uncertainty in the final atomic state. The atom and the field must then be described by density operators instead of pure states. The formula (1) remains valid with $P(j,\phi|n)$ replaced by $P^{(\exp)}(j,\phi|n)$. This is justified by Bayes' law or by generalization of the measurement postulate to statistical mixtures[33].

In order to obtain more information, we repeat the process and send a sequence of atoms across C. This results in a step-by-step change of the photon number distribution. From one atom to the next, we vary $\phi$. Calling $\phi(k)$ the detection angle for the $k$th atom and $j(k)$ its spin reading, the photon number distribution after $N$ atoms becomes:

$$P_N(n) = \frac{P_0(n)}{Z} \prod_{k=1}^{N} \left[ A + B\cos(n\Phi - \phi(k) + j(k)\pi) \right] \quad (2)$$

where $Z$ enforces normalization. For an efficient decimation, we alternate between detection directions nearly coinciding with the vectors associated with $q$ non-orthogonal $|+_p\rangle$ states. Each atom has a chance to reduce the probability of a photon number different from the one decimated by its predecessor. After a finite number of steps, numerical simulations predict that a single $n$ value (modulo $2q$) survives.

## Observing the field-state collapse

We have applied this procedure to a coherent microwave field at 51.1 GHz stored in an ultrahigh-$Q$ Fabry–Pérot cavity made of niobium-coated superconducting mirrors[28]. Our set-up is described in ref. 29. The cavity has a very long damping time $T_c = 0.130$ s. It is cooled to 0.8 K (average thermal photon number $n_t = 0.05$). The field is prepared by coupling a short microwave pulse into C (by way of diffraction on the mirrors' edges[28]). Its photon number distribution and average photon number, $n_0 = 3.82 \pm 0.04$, are inferred from the experimental data (see below). Our single-photon-sensitive spin-clocks are circular Rydberg atoms of rubidium. They cross C successively, separated on average by $2.33 \times 10^{-4}$ s. Parameters are adjusted to realize a $\sim\pi/4$ clock shift per photon (Methods), corresponding to eight positions of the spin on the Bloch sphere (Fig. 1b). This configuration is adapted to count photon numbers between 0 and 7. For $n_0 = 3.82$, the probability for $n \geq 8$ is 3.5%.

Four phases $\phi_i$ ($i = a, b, c, d$), corresponding to directions pointing approximately along the spin states associated with $n = 6, 7, 0, 1$, are used, in random order, for successive atoms (Methods). A sequence of $j$ values can be decoded only when combined with the corresponding phase choices, in analogy with the detection basis reconciliation of quantum key distribution protocols[36]. Figure 2a shows the data from the first 50 detected atoms, presented as ($j$, $i$) doublets, for two independent detection sequences performed on the same initial field.

From these real data, we compute the products of functions $\Pi_N(n) = \prod_{(k=1...N)} [A + B\cos(n\Phi - \phi_{i(k)} + j(k)\pi)]$. The $A$, $B$, $\Phi$ and $\phi_i$ values are given by Ramsey interferometer calibration (Methods). The evolutions of $\Pi_N(n)$, displayed as functions of $n$ treated as a continuous variable, are shown in Fig. 2b for $N$ increasing from 1 to 50. The $\Pi_N(n)$ functions converge into narrow distributions whose widths decrease as more information is acquired. These functions are determined uniquely by the experimental data. Their evolution is independent of any a priori knowledge of the initial photon distribution. The data sequence itself, however, depends of the unknown state of the field, which the measurement reveals.

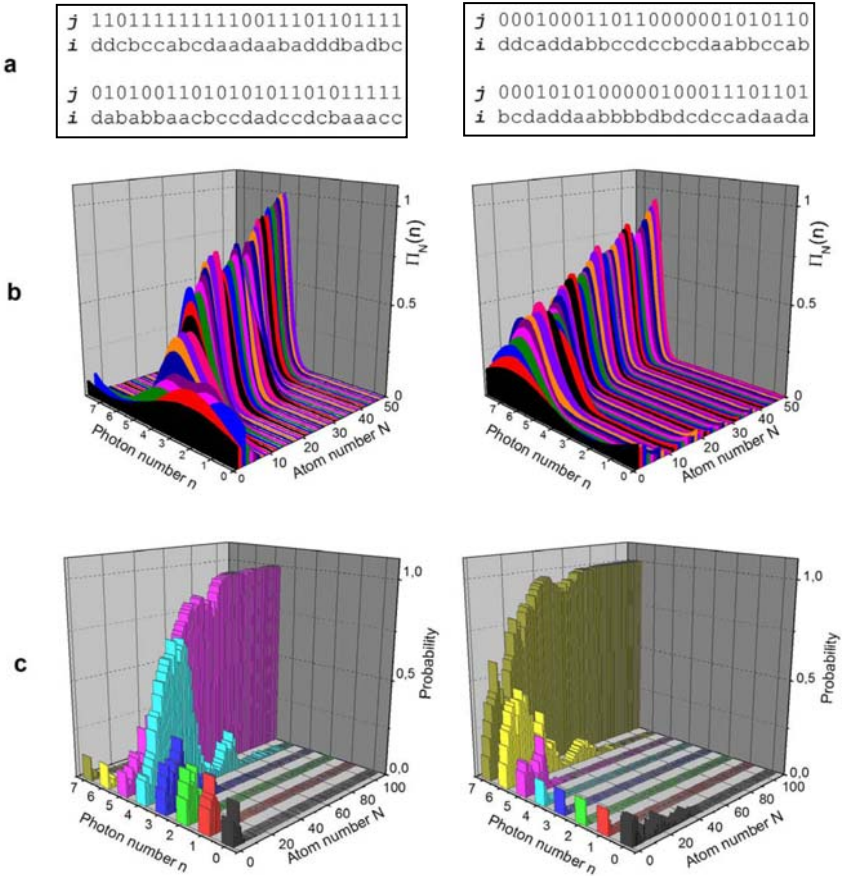

**Figure 2 Progressive collapse of field into photon number state. a**, Sequences of ($j$, $i$) data (first 50 atoms) produced by two independent measurements. **b**, Evolution of $\Pi_N(n)$ for the two sequences displayed in **a**, when $N$ increases from 1 to 50, $n$ being treated as a continuous variable (integral of $\Pi_N(n)$ normalized to unity). **c**, Photon number probabilities plotted versus photon and atom numbers $n$ and $N$. The histograms evolve, as $N$ increases from 0 to 110, from a flat distribution into $n = 5$ and $n = 7$ peaks.

Inserting $\Pi_N(n)$ into equation (2) and extending the procedure to $N = 110$, we obtain the evolution of the photon number histograms for these two realizations (Fig. 2c). These histograms show how our knowledge of the field state evolves in a single measuring sequence, as inferred from baysian logic. The initial distributions ($P_0(n) = 1/8$) are flat because the only knowledge assumed at the beginning of each sequence is the maximum photon number $n_{max}$. Data are analysed after the experiment, but $P_N(n)$ could also be obtained in real time. The progressive collapse of the field into a Fock state (here $|n = 5\rangle$ or $|n = 7\rangle$) is clearly visible. Information extracted from the first 20 to 30 atoms leaves an ambiguity between two competing Fock states. After ~50 atoms (detected within ~0.012 s), each distribution has turned into a main peak with a small satellite, which becomes totally negligible at the end of the two sequences.

## Reconstructing photon number statistics

Repeatedly preparing the field in the same coherent state, we have analysed 2,000 independent sequences, each made of 110 ($j$, $i$) doublets recorded within $T_m \approx 0.026$ s. This measuring time is a compromise. Short compared to $T_c$, it is long enough to allow for a good convergence of the photon number distribution. We have computed the mean photon number $\langle n \rangle = \sum_n n P_N(n)$ at the end of each sequence. The histogram of these $\langle n \rangle$ values, sampled in bins of width 0.2, is shown in Fig. 3. Peaks at integers appear on top of a small background due to sequences that have not fully collapsed, or that have been interrupted by field decay.

The histogram of the peaks in Fig. 3 directly reveals the photon number probability distribution of the initial field, modified by damping during the measurement. Disregarding the 23% background, we fit the experimental histogram of integer values to a Poisson law with $\langle n \rangle_{ave} = 3.46 \pm 0.04$ (blue circles), and normalized to 0.77 (probability of fully converged sequences). This is the expected distribution for a coherent field with an initial mean $n_0 = 3.82 \pm 0.04$, after decay during the time $T_m/2 \approx T_c/10$. Remarkably, the non-converged sequences do not introduce any noticeable bias in the distribution of fully collapsed measurements. The experimental excess probability of $0.019 \pm 0.006$ for $n = 0$ is well understood. It is due to the measurement being performed modulo 8, which attributes $n = 8$ events (0.012 probability) to the $n = 0$ bin. The near-perfect agreement of the fit with the experiment provides a direct verification of the quantum postulate about the probabilities of measurement outcomes.

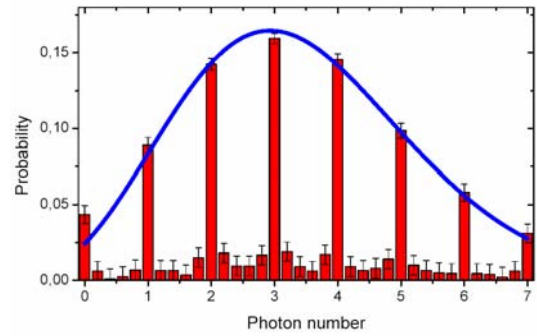

**Figure 3 Reconstructed photon number distribution.** Histogram of $\langle n \rangle$ values obtained from 2,000 QND collapse sequences (each involving $N = 110$ atoms). The $\langle n \rangle$s are sampled in intervals of 0.2. The error bars are the statistical standard deviations. The peaks at integer numbers reveal Fock states. The background is due to incomplete or interrupted collapses. Data shown as blue circles are obtained by fitting the distribution of integer number peaks to a Poisson law, yielding $\langle n \rangle_{ave} = 3.46 \pm 0.04$ (the blue line represents a continuous Poisson distribution joining the circles as a guide for the eye).

## Repeated measurements and field jumps

Repeatability is another fundamental feature of an ideal QND measurement. To test it, we follow the evolution of the field state along sequences made of ~2,900 atoms. We determine $P_N(n)$ and $\langle n \rangle$ up to $N = 110$. We then drop the first atom and replace it with the 111st one, resuming the calculation with a flat initial distribution and obtain a new $\langle n \rangle$. We repeat the procedure atom by atom. We thus decode continuously a single field history versus time. Measurements separated by more than $T_m$ exploit independent information.

Figure 4a shows the evolution of $\langle n \rangle$ over 0.7 s for the two sequences whose initial data are displayed in Fig. 2a. In each case, $\langle n \rangle$ evolves quickly towards an integer (5 or 7). This collapse is

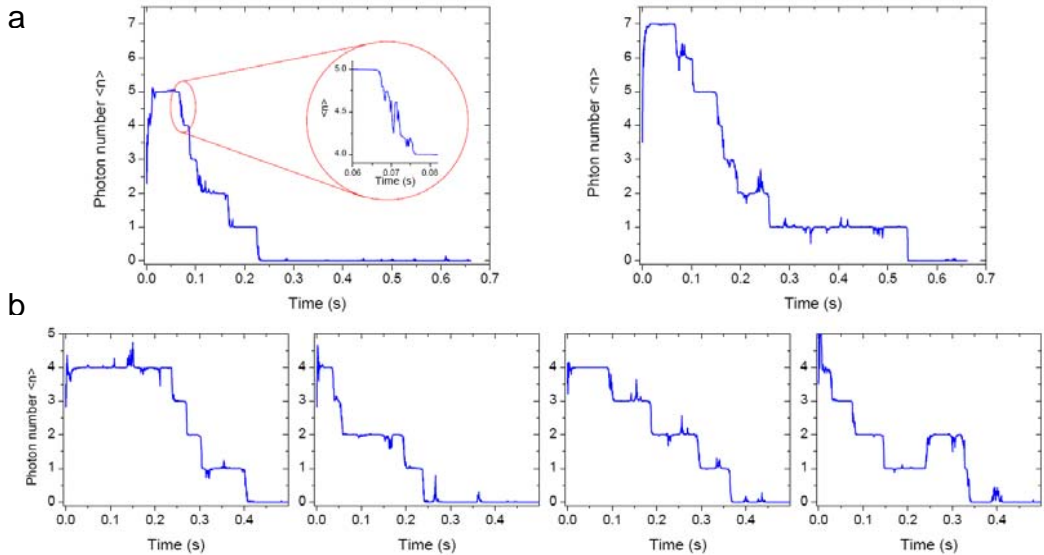

**Figure 4 Repeated QND measurements. a**, Mean photon number $\langle n \rangle$ followed over 0.7 s for the two sequences whose collapse is analysed in Fig. 2. After converging, $\langle n \rangle$ remains steady for a while, before successive quantum jumps bring it down to vacuum. Inset, zoom into the $n = 5$ to 4 jump, showing that it is detected in a time of ~0.01 s. **b**, Four other signals recording the evolution of $\langle n \rangle$ after field collapse into $n = 4$. Note in the leftmost frame the exceptionally long-lived $n = 4$ state, and in the rightmost frame the $n = 1$ to 2 jump revealing a thermal field fluctuation.

followed by a plateau, corresponding in these two realizations to ~2 independent measurements. Eventually, cavity damping results in a photon loss: a quantum jump occurs, decreasing $\langle n \rangle$ by one. This event is recorded after a delay of a fraction of $T_m$ with respect to the real jump time, as several atoms are required after the quantum leap to build up the new photon number probability. Note in the inset of Fig. 4a that it takes about 0.01 s for the atoms to 'realize' that a jump has occurred. The staircase-like evolution of the field proceeds in this way down to vacuum. Figure 4b presents four other examples of signals following a collapse into $n = 4$. The randomness of the step durations is typical of quantum dynamics. In one of these recordings (leftmost panel), the $n = 4$ step lasts 0.235 s, corresponding to ~9 independent QND measurements

The lifetime of a $n$-Fock state, $T_c/n$ at $T = 0$ K (ref. 39), is reduced by thermal effects to $T_c/[n + n_t(2n + 1)]$, that is, ~0.029 s, ~0.023 s and ~0.017 s for $n = 4$, 5 and 7, respectively. The statistical analysis of 2,000 QND sequences, each of 0.7 s duration, provides a detailed description of the dynamical evolution under cavity relaxation of Fock states with $n$ up to 7 (J.B. *et al.*, manuscript in preparation). The sequences of Fig. 4, in which some Fock states survive much longer than their lifetimes, are relatively rare events. We have selected them to demonstrate the ability of the QND procedure to generate and repeatedly measure large $n$-Fock states of radiation in a cavity.

### Beyond energy measurements

Our QND source of Fock states operates in a way different from previous methods based on resonant[25,40] or Raman[41] processes in cavity quantum electrodynamics, which have so far been limited to smaller photon numbers ($n = 1$ or 2). This QND measurement opens novel perspectives for the generation of non-classical states of light. If the initial photon number distribution spans a range of $n$ values larger than $2q$, the decimations induced by successive atoms do not distinguish between $n$ and $n + 2q$. The field then collapses into a superposition of the form $\sum_q c_{n+2q}|n + 2q\rangle$. For instance, $c_0|0\rangle + c_{2q}|2q\rangle$ represents a field coherently suspended between vacuum and $2q$ photons. This superposition of states with energies differing by many quanta is a new kind of 'Schrödinger cat' state of light.

Superpositions of field states with the same amplitude but different phases—'Schrödinger cats' of a different kind—are also generated in this experiment. As the photon number converges, its conjugate variable, the field's phase, gets blurred. After the first atom's detection, the initial state collapses into a superposition of two coherent states with different phases[31,42]. Each of these components is again split into two coherent states by the next atom and so on, leading to complete phase-uncertainty when the photon number has converged[31]. The evolution of the Schrödinger cat states generated in the first steps of this process could be studied by measuring the field Wigner function[43]. Decoherence of superpositions of coherent states[44,45] containing many photons could be monitored in this way.

### METHODS SUMMARY

The preparation and detection of circular Rydberg atoms, the cavity and the Ramsey interferometer are described elsewhere[20,26,29]. The $|\pm_z\rangle$ states are the circular Rydberg levels of rubidium with principal quantum numbers 51 and 50 (transition frequency ~51.1 GHz). The theoretical phase shift per photon[20,26] is $\Omega^2 t/2\delta$, where $\Omega/2\pi = 50$ kHz is the vacuum Rabi frequency at cavity centre, $\delta/2\pi$ is the atom–cavity detuning and $t = 3 \times 10^{-5}$ s is the effective atom–cavity interaction time. It is defined as $t = (\pi/2)^{1/2}w/v$, where $w$ is the waist of the gaussian cavity field mode ($w = 6$ mm) and $v$ the atomic velocity ($v = 250$ m s$^{-1}$). This effective time is obtained by averaging the spatial variation of the square of the atom–field coupling as the atom crosses the cavity mode[20,26].

More details about the experimental settings, including the determination of the $A$ and $B$ parameters, the fine tuning of the phase shift per photon, $\Phi$, and the adjustment of the four phases of the Ramsey interferometer are given in the Methods section. We also analyse the adiabaticity of the atom–field coupling, which is an essential feature of our measurement. We describe the generation of sequences of atoms crossing the cavity one at a time with a well-defined velocity, and we discuss the effect of rare multi-atom events on the detection signals. We also explain why the sequences of detection directions $\phi_i$ occur randomly in a measuring sequence. We conclude by discussing alternative strategies to pin down the photon number non-destructively.

## METHODS

### Experimental settings

To calibrate the Ramsey interferometer, we send atoms prepared in $|-_z\rangle$ across the set-up, with C empty. We measure the probability $P(j = 0, \phi)$ for detecting the atom in $|+_z\rangle$ as a function of the relative phase $\phi$ between $R_1$ and $R_2$. From the Ramsey fringes, we obtain the phase origin $\phi = 0$ corresponding to a detection along the $|+_0\rangle$ spin direction.

Tuning of $\delta$ is performed by moving the cavity mirrors with piezoelectric actuators. In theory, a $\pi/4$ phase shift corresponds to $\delta/2\pi = 300$ kHz. We set the detuning close to this value (with a 15 kHz uncertainty due to imperfect knowledge of the atomic transition frequency affected by small residual Stark and Zeeman shifts). With this detuning, the photon-induced phase shift is, within 3%, a linear function of $n$ for $n = 0$ to 7.

By Stark-shifting the atomic spin phase with a short electric field pulse applied just before $R_2$ and adjusted to different amplitudes, we translate the fringes and set $\phi$ to four different values, close to $l\pi/4$ ($l = -2, -1, 0$ and 1). After preliminary settings of $\Phi$ and $\phi_i$, we refine our calibration. We inject a small coherent field in C ($n_0 = 1.2$). This field has a negligible probability for $n > 4$. For each $\phi_i$, we record the fraction $\eta_0(\phi_i)$ of spins found in $|+_z\rangle$ on a sequence of atoms crossing C in a time short compared to $T_c$. Repeating the sequence many times, we find distributions of $\eta_0(\phi_i)$s, which we fit as a sum of five peaks centred on the discrete values equal to $P^{(\text{exp})}(j = 0, \phi_i|n) = [A + B\cos(n\Phi - \phi_i)]/2$, with $n = 0$ to 4. From a best fit of the $\eta_0$ distributions, we get the values $A = 0.907 \pm 0.004$, $B = 0.674 \pm 0.004$, $\Phi/\pi = 0.233 \pm 0.004$, $\phi_i/\pi = -0.464 \pm 0.013$, $-0.229 \pm 0.009$, $-0.015 \pm 0.007$ and $+0.261 \pm 0.006$ ($i = a, b, c, d$). These values are inserted in equation (2). The fringe contrast $B$ is reduced below 1 by experimental imperfections (stray fields, detection errors, two-atom events, see below). The other six parameters $A$, $\Phi/\pi$ and $\phi_i/\pi$ ($i = a, b, c, d$) are close to their ideal values (1, ¼, −½, −¼, 0, ¼, respectively).

## Adiabaticity

The adiabatic variation of the coupling as the atoms cross the gaussian profile of the cavity mode keeps the atomic emission rate extremely low. The theoretical probability that an atom deposits an additional photon when the cavity contains a coherent field with $n_0 = 3.82$ is below $1.3 \times 10^{-6}$. Consistent with this very small value, we found on analysing our experimental data that the average number of photons deposited in C by a sequence of ~2,900 atoms is negligible compared to $n_t$.

## Atomic sequences

They are realized by pulsing the Rydberg atom preparation. The atoms are excited from a thermal atomic beam, velocity selected by optical pumping, at a rate of $1.4 \times 10^4$ pulses per second. The velocity spread, $\Delta v = \pm 1$ m s$^{-1}$ around 250 m s$^{-1}$, has a negligible effect on the Ramsey fringe contrast. In order to limit the number of events with two atoms per pulse, the intensity of the exciting lasers is kept low (average number of detected Rydberg atoms per pulse is 0.3, detection efficiency 50%). Undetected atoms do not affect the photon number distribution. A single atom is counted in 22% of the preparation pulses, while 3% of them contain a detected atom pair. When two atoms (whether detected or not) cross C together there is a slight reduction of the interferometer contrast, owing to small cavity-mediated interactions[46] between the atoms. This reduction is taken into account in the measured $B$ value. All detected events with one or two atoms per pulse are compiled independently in the data analysis. When three atoms are in C together the fringe contrast is strongly reduced, but the probability of these events is small (2% probability for preparing 3 atoms or more per pulse).

## Randomness of detection directions

The interferometer phase is changed from pulse to pulse, going cyclically from $\phi_a$ to $\phi_d$. As the presence of one (or two) atoms in a given pulse is random, we cannot predict which phase will correspond to the next observed atom. We acquire this knowledge by detecting the atom, and $i(k)$ is thus a randomly measured variable.

## Other QND measurement strategies

Efficient photon number decimation could be obtained by alternative methods. As suggested in refs 30 and 31, we could change the atom–cavity interaction time (and hence the phase shift per photon) by detecting randomly atoms from a thermal atomic beam, without velocity selection. The optimal data acquisition procedure consists in applying to successive atoms a sequence of $\pi, \pi/2, \pi/4 \ldots$ phase shifts per photon, while adjusting $\phi$ for each spin, based on the result of the previous measurement. This expresses $n$ in binary code, each atom providing a bit of information[20,47]. The required number of atoms per measuring sequence is then minimal, equal to the smallest integer $\geq \log_2(n_{max} + 1)$. This ideal strategy requires however a deterministic beam of atoms, with perfect Ramsey fringe contrast and 100% detection efficiency.

**Acknowledgements** This work was supported by the Agence Nationale pour la Recherche (ANR), by the Japan Science and Technology Agency (JST), and by the EU under the IP projects SCALA and CONQUEST. C.G. and S.D. were funded by the Délégation Générale à l'Armement (DGA). J.M.R. is a member of the Institut Universitaire de France (IUF).

**Author Information**: Correspondence should be addressed to M.B. (brune@lkb.ens.fr) or S.H. (haroche@lkb.ens.fr).